\documentclass[a4paper]{jpconf}
\usepackage{graphicx}
\usepackage{natbib}

\def\arcsec{\nobreak{$''$}}
\def\spose#1{\hbox to 0pt{#1\hss}}
\def\simlt{\mathrel{\spose{\lower 3pt\hbox{$\mathchar"218$}}
     \raise 2.0pt\hbox{$\mathchar"13C$}}}
\def\simgt{\mathrel{\spose{\lower 3pt\hbox{$\mathchar"218$}}
     \raise 2.0pt\hbox{$\mathchar"13E$}}}

\begin{document}
\title{Do stellar winds play a decisive role in feeding AGN?}

\author{R. Davies, L. Burtscher, K. Dodds-Eden, G. Orban de Xivry}

\address{Max-Planck-Institut f\"ur extraterrestrische Physik, \\Postfach 1312, Giessenbachstr., 85741 Garching, Germany}

\ead{davies@mpe.mpg.de}

\begin{abstract}
While the existence of a starburst-AGN connection is undisputed, there is no consensus on what the connection is.
In this contribution, we begin by noting that the mechanisms which drive gas inwards in disk galaxies are generally inefficient at removing angular momentum, leading to stalled inflows.
Thus, a tiered series of such processes is required to bring gas to the smallest scales, each of which on its own may not correlate with the presence of an AGN.
Similarly, each may be associated with a starburst event, making it important to discriminate between `circumnuclear' and `nuclear' star formation.
In this contribution, we show that stellar feedback on scales of tens of parsecs plays a critical role in first hindering and then helping accretion.
We argue that it is only after the initial turbulent phases of a starburst that gas from slow stellar winds can accrete efficiently to smaller scales. 
This would imply that the properties of the obscuring torus are directly coupled to star formation and that the torus must be a complex dynamical entity.
We finish by remarking on other contexts where similar processes appear to be at work.
\end{abstract}

\section{Introduction}
\label{sec:intro}

The starburst-AGN connection has been a long disputed topic.
A large part of the reason is that any conclusion will depend on the spatial and temporal scales being considered, as well as the luminosities of the AGN and the properties of their host galaxies.
The other important issue is whether one is assessing a `casual' or `causal' relation (both of which play a role in our understanding of galaxy evolution and the starburst-AGN connection).
Casual connections -- in which star formation and an AGN are independent products of another phenomenon -- abound.
Causal connections -- in which either the star formation or AGN has a direct impact on the other -- are more elusive, and form the rationale for the work discussed here.
In this contribution we focus on local galaxies, and hence on Seyferts and Low Luminosity AGN (LLAGN).
In addition, we consider primarily AGN with spiral hosts, since local Seyferts typically reside in disk galaxies (although the situation changes if one includes LINERs; \citealt{ho08}).
Our aim is to touch on a number of issues related to the starburst-AGN connection in such galaxies, starting with the largest scales ($\simgt1$\,kpc) in Section~\ref{sec:largescales}, moving in towards the 0.1--1\,kpc scale circumnuclear region in Section~\ref{sec:circumnuclear}, and then the central tens of parsecs in Sections~\ref{sec:starburst} and~\ref{sec:winds}.
We argue that on these small scales there is indeed a causal relation between star formation and the AGN -- with stellar outflows from a young post-starburst first hindering and then helping gas inflow towards the central supermassive black hole (BH).

\section{Are large scales important?}
\label{sec:largescales}

It is easy to forget that the requirements for fuelling star formation (i.e. cold, dense gas) and a central BH (i.e. low angular momentum gas) are different.
That star formation requires cold, dense gas has recently been beautifully demonstrated by \cite{lad10}. 
They showed that the number of young stellar objects in galactic clouds is not related to the total cloud mass, but to the mass above a threshold K-band extinction of $A_K=0.8$\,mag.
One can argue that this should lead to an anti-correlation between star formation and AGN: only if inflow is stalled (ultimately resulting in a reduced accretion rate onto the central BH) will the gas pile up enough to form the cold dense cores that lead to star formation.

In contrast, \cite{jog06} emphasized that even gas located at a radius of a few hundred parsecs has to lose 99.99\% of its angular momentum before it reaches the central BH.
This means that large scale structures in a galaxy can only be related to AGN activity if they remove sufficient angular momentum.
A galaxy interaction is such a process, and leads to the infall of a huge gas reservoir to small spatial scales on a short timescale.
It is for this reason that the most luminous AGN tend to be associated with mergers.
However, it is also important to realise that this infall is unlikely to feed the BH(s) directly, but only bring the gas down to scales of order 100\,pc.
From there other mechanisms, perhaps related to disk processes as suggested by the nested simulations of \cite{hop10}, will drive the gas to smaller radii.
For Seyferts in disk galaxies, the efficiency of angular momentum removal is the reason why there is at best only a marginal relation between the presence of large scale bars and AGN \citep{lai02,lau04}.
Bars are inefficient because they only drive modest inflows, and what there is always stalls at a ring on scales larger than 100\,pc \citep{reg04}.
In these cases, it is processes acting on smaller scales that will predominantly regulate the rate at which gas can reach the central regions.

It is in this context that Narrow Line Seyfert 1 (NLS1) galaxies are so unusual.
They provide an example of how, in a specific subset of AGN, the large scale host galaxy properties play a prominent role in the properties of the central BH.
Combining several analyses from the literature, \cite{orb11} showed that most NLS1s are associated not only with strong bars, but also with circumnuclear grand design dust spirals.
In these objects, it seems likely that circumnuclear spiral pattern is a standing wave driven by the bar.
Indeed, this is the way that such spirals are produced in hydrodynamical simulations \citep{mac04}.
Thus, the AGN activity is associated with the bar, because the bar stimulates a circumnuclear spiral to drive gas to smaller scales than it otherwise could.
Thus the strong bars in NLS1s can be considered efficient at removing angular momentum.
\cite{orb11} also measured the S\'ersic index of the bulge luminosity profiles of NLS1s, together with a sample of Broad Line Seyfert~1s.
They found that while the indices for the BLS1s covered the full range expected for classical and pseudobulges, the NLS1s were all clustered at the low end.
The implication is that these galaxies have pure pseudobulges, and have therefore grown and evolved primarily through secular processes.
The authors speculated that in this case, because the BHs have grown through accretion of disk material, they should be spinning rapidly.
This would make them radiatively efficient, which in turn slows their growth.
If this scenario is confirmed, it could provide a natural explanation for the high BH spins observed in the majority of NLS1s via the 6.4\,keV Fe~K$\alpha$ line, as well as the typically low masses of BHs in all galaxies with pure pseudobulges.

\section{Circumnuclear Inflow}
\label{sec:circumnuclear}

It has been known now for a long time that the circumnuclear region (scales of 100--1000\,pc) plays a crucial role in regulating gas inflow, but the complexity of its dynamics has only begun to be appreciated over the last decade \citep{wad04}.
Observationally, the first clues came from imaging surveys with {\em HST} of dust structures, which are presumed to be associated with gas and hence the inflow mechanisms.
In matched samples of active and inactive galaxies, \cite{mar03} found a variety of dust structures but no correlation with the presence of an AGN.
The first hints of such a relation were reported by \cite{sim07} who, by separating AGN according to host galaxy Hubble type, showed that while circumnuclear dust structures are necessary for AGN activity, they are not sufficient, and that other factors must also be important.

Taking a different approach, the NUGA team looked at the molecular gas directly using 1--2\arcsec\ resolution CO data.
They calculated the gravitational torques, to assess whether these could drive the gas into the central $\sim100$\,pc. 
Intriguingly, of the 7 galaxies analysed by \cite{haa09}, only 4 had a significant global negative torque driving a net inflow through the circumnuclear region; and \cite{gar11} reported that of the 25 objects in the full study, only 1/3 revealed torques that could drive gas into the central 100\,pc.
It is possible that this perplexing result can be understood if one takes account of the AGN luminosity, which is typically a hidden parameter.
The NUGA sample was designed to span a range of AGN types which means that a significant number are LINERs.
And it contains a large number of LLAGN:
none of the objects in \cite{haa09} appears among the AGN detected at 14--195\,keV in the {\em Swift} BAT 58-month catalogue \citep{bau10}; and the 2--10\,keV luminosities (where available) are all $<10^{40}$\,erg\,s$^{-1}$.
We speculate that the very modest inflow needed to fuel a LLAGN -- less than $10^{-3}$\,M$_\odot$\,yr$^{-1}$ -- is so small that it does not require, and perhaps may even tend to preclude, a coherent dynamical process in the circumnuclear region.
In contrast, such processes may typically help supply the larger 0.01--0.1\,M$_\odot$\,yr$^{-1}$ inflow rates of more luminous Seyferts.
We argue that to advance our understanding of the role of the circumnuclear region in fuelling AGN, it is necessary to assess the mechanisms and inflow rates on these scales as a function of AGN luminosity.

None of the work described above directly uses the observed kinematics.
Such analyses can be difficult and require high signal-to-noise data that traces kinematics at high spatial resolution across a full 2D field.
This is because the velocity fields are usually dominated by rotation, and it is only the deviations from this -- the residuals -- that contain information about the inflow.
\cite{haa09} performed a harmonic analysis on the CO kinematics for their sample, a technique that was also used by \cite{van10} on the ionised gas kinematics in NGC\,1097 (and is also the basis of kinemetry \citep{kra06} which was used in the analysis of the SAURON sample of early type galaxies).
An alternative is to fit a disk model to the circular motions and assess the residuals, a technique employed by a number of authors \citep[e.g.][amongst others]{sto07,rif08,dav09,sch11}.
While this can be effective at revealing non-circular motions, interpreting them robustly can be hard.
For example, \cite{dav09} have pointed out that the line-of-sight kinematic signature of inflow along a 2-arm spiral is a 1-arm kinematic spiral (superimposed on the global rotation) -- making it challenging to relate the kinematics to inflow along spiral arms in a specific way.
While the case studies performed so far have demonstrated the potential of such work, systematic studies using integral field spectroscopy (e.g. Hicks et al. in prep) or millimetre line interferometry to probe the kinematics and their residuals will be needed to assess the variety of mechanisms that may be at work in the circumnuclear region, and how they relate to AGN activity in different luminosity regimes.

\section{Young Nuclear Starbursts}
\label{sec:starburst}

The mechanisms discussed in the preceeding sections are inefficient at removing angular momentum.
As a result, the inflows they drive are stalled in the central regions, leading to intense periodic bursts of star formation.
The key issue we want to address here is whether these young nuclear starbursts have an impact on AGN fuelling or is simply a serendipitous by-product of a stalled inflow.
One can only begin to answer this by first clarifying what we mean by `nuclear' and `young'.

Historically, all starbursts that occur in the central kiloparsecs of a galaxy have been called `nuclear'.
Now that we can spatially resolve this region, such nomenclature is inappropriate, and it is instead crucial to discriminate between `circumnuclear' meaning 0.1--2\,kpc and `nuclear' meaning $\simlt$50\,pc (and in the future, even smaller scales).
The separation of these scales, although to some extent arbitrary, is based on both observational and theoretical work.
\cite{com10} presented a survey of 113 star-forming rings in the central regions of 107 galaxies, finding that such rings occur in 20\% of all disk galaxies.
The radii of these rings spans 0.1--2\,kpc, 
matching the range in the hydrodynamical simulations of \cite{reg03}.
These authors found that rings migrate inwards from $\sim1$\,kpc as they accumulate gas, and can reach sizes as small as $\sim0.2$\,kpc.
It is a small step to realise that the frequency of rings should be greater if one includes those that are not currently forming stars (i.e. without H$\alpha$ emission).
Indeed, molecular rings are commonly found in interferometric millimetre observations such as those from the PdBI -- e.g. at a radius of 140\,pc (1.7\arcsec) in NGC\,3227 \citep{dav06} and at 500\,pc (5\arcsec) in NGC\,6951 \citep{kri07}.
And spatially resolved stellar population synthesis has revealed intermediate age ($\sim0.5$\,Gyr) rings at radii of 600\,pc (1.5\arcsec) in Mkn\,1066 \citep{rif10} and 300\,pc (1\arcsec) in Mkn\,1157 \citep{rif11}.

That rings are indeed common, and may be rather small, must be borne in mind when interpreting observations where spatial resolution is insufficient to disentangle `circumnuclear' and `nuclear'.
One example of this is the analysis of the 3.3\, $\mu$m PAH emission feature (which traces star formation) in type~1 and~2 Seyferts by \cite{ima04}.
At the typical redshift of their sample, the spectroscopic aperture covered $\sim0.8$\,kpc.
Thus most of the PAH detections in Seyfert~2s and as many as half of those in Seyfert~1s may be attributable to circumnuclear star formation.
Such a hypothesis would have to be verified by higher resolution observations, but it would imply that the star formation results from stalled inflows, at radii too large to have a direct causal impact on BH accretion.

The second issue here is to ask what we mean by `young' and whether such populations are associated with AGN.
The largest survey to address this made use of stellar population synthesis for the central $\sim200$\,pc of 65 Seyfert~2 (and 14 other) galaxies \citep{cid04}.
By fitting the absorption features in optical spectra with a library of model templates, the authors derived the star formation history of their targets. 
Most of the sample showed evidence for a 20--60\% contribution from what could be either a young ($<$25\,Myr) population or a featureless continuum.
But the higher the contribution of this component, the stronger were the weak broad emission lines, suggesting that it was associated with the AGN.
The authors noted that while in some objects there may be a young dusty starburst, its contribution to the optical continuum is $\simlt20$\%. 
Their conclusion was simply that ``1/3 to 1/2 of Seyfert~2s have experienced significant star formation in the recent past'' (specifically, 40\% of their galaxies have a mean age in the central $\sim200$\,pc less than 300\,Myr).
\cite{rif09} found a similar result from near infrared spectral synthesis of the central $\sim300$\,pc of 24 Seyferts.
In this case, a significant contribution from either the young stellar population or featureless continuum was found in 50\% of the Seyfert~2s and most of the Seyfert~1s. 
This is what one would qualitatively expect if the component was associated with the AGN.
Thus while a young ($<50$\,Myr) population may be present in some objects, it is not possible to quantify definitively its scale or location.
In an analysis of STIS spectra of 22 Seyfert~2s (with an aperture radius of $\sim30$\,pc, thus avoiding issues associated with circumnuclear rings), \cite{sar07} found that nearly half of the objects required a young starburst or featureless continuum, and most of the rest had only old populations.
They argued that the young starburst interpretation of this component contradicted the dearth of intermediate age populations, and that instead its strong correlation with high excitation lines argues for its association with the AGN.

There seems to be some disagreement in the studies above about whether AGN are associated with intermediate age (0.1--1\,Gyr) stars; and it seems that some, but probably not all, do host such a population.
But a common conclusion appears to be that AGN are not obviously associated with young ($<50$\,Myr) starbursts in the nuclear ($<50$\,pc) region.


\section{The Role of Stellar Outflows}
\label{sec:winds}

Using adaptive optics to achieve resolutions of order $\sim10$\,pc, \cite{dav07} estimated the characteristic age of the most recent burst of star formation in a number of AGN and compared this to the accretion rate.
They found ages of $<50$\,Myr were associated with lower Eddington ratios, and the accretion rate was only high for ages $>50$--100\,Myr.
They suggested that this was because at very young ages, the outflows from stars (OB winds and type~II supernovae) injected so much turbulence into the surrounding inter-stellar medium (ISM), that any gas would tend to be ejected rather than accreted.
However, for a short burst of star formation, this phase would end after 50--100\,Myr, after which the slower outflows from AGB stars, which would remain gravitationally bound to the cluster, would be available to accrete to smaller scales.
Based on these arguments, \cite{vol08} put together a scenario in which the star formation and torus properties would be time-dependent, driven by (the stochastic nature of) changes in gas flow from larger scales.
At about the same time, 3D hydrodynamical simulations, subsequently scaled to match a typical Seyfert, were yielding similar conclusions about stellar feedback \citep{sch09,sch10}.
These showed that for a stellar cluster of $10^8$\,M$_\odot$ in the tens of parsecs around a $10^7$\,M$_\odot$ BH, outflows with speeds greater than about 200\,km\,s$^{-1}$ would tend to disrupt the ISM and drive gas out; but slower outflows could be accreted in streamers to smaller scales on a timescale between the end of the type~II supernova phase and before type~Ia supernova begin to appear.
In this particular case, scaled to match NGC\,1068, the streamers fed a dense and compact ($10^6$\,M$_\odot$, 0.5--1\,pc) turbulent disk.
The size of this disk is set by the `angular momentum barrier' and depends on the kinematic properties of the surrounding stellar cluster.
Remarkably, it matched the size of the maser disk in this object and also the size of the compact mid-infrared continuum source detected with VLTI.
Although this is so far only one example, it provides a tantalising suggestion that we may be able to directly link the properties of a star forming cluster on scales of 10--50\,pc with those of more compact structures on scales of 0.1--1\,pc.
It is reasonable to assume that these structures are related to the putative obscuring torus.
This leads immediately to two important conclusions:
(i) that the torus is complex, comprising structures on at least 2 different scales that may be responsible for different observational aspects \citep[see also][]{eli08,elv12};
(ii) the torus is a dynamical entity, the time-dependent properties of which are related to star formation in its local environment, and ultimately to gas inflow from larger scales.

It is important to realise that other models are possible.
The analytical model of \cite{kaw08} and detailed 3D hydrodynamical simulations of \cite{wad09} suggest that inflow and accretion onto the BH should increase with the turbulence in the ISM, and hence occur simultaneously to the star formation.
\cite{hob11} suggest that supernovae should be able to drive enough of the ambient ISM inwards in a `ballistic' mode to fuel an AGN.
All these scenarios rely on the role played by stellar outflows in driving gas to smaller scales.
The difference between them is whether gas inflow is synchronous with, or subsequent to, the star formation episode.
Further testing with detailed observations and simulations for more objects will be needed to discriminate between them.
It is, however, notable that the same physical processes as proposed by \cite{dav07} and \cite{sch09,sch10} have been invoked to explain similar situations in a number of other environments.
These include the inner 0.5\,pc of the Galactic Center, where simulations have shown that fast stellar winds, and the angular momentum of their stars from which they come, are inhibiting accretion to smaller scales \citep{cua06,cua08}.
In the central 2\,pc of M\,31, \cite{cha07} have argued that mass loss from the older red stars in the eccentric stellar disk (nuclei P1 and P2 in the usual notation) can be driven to smaller scales where it is able to produce a compact gravitationally unstable gas disk of $\sim10^5$\,M$_\odot$ every 0.1--1\,Gyr. 
It can thus fuel the star formation that has been observed in the smaller (0.2\,pc scale) younger (200\,Myr) cluster P3.
It is also believed that the multiple stellar populations of globular clusters may arise from secondary generations of stars being formed from the outflows of earlier generations.
\cite{der08} have shown that there is a specific window around an age of 40\,Myr -- between the phases of type~II and type~Ia supernovae -- when this can happen.

\section{Conclusions}
\label{sec:conc}

The main points raised in this contribution can be summarised as follows:
\begin{itemize}
\item
Global trends between AGN and large scale phenomena are masked by the variety and number of mechanisms required to bring gas to small scales; but relations may be found in specific subsets of objects (e.g. NLS1s).
The AGN luminosity -- as a proxy for accretion rate -- may be an important (hidden) parameter when searching for such relations, since different phenomena will lead to different gas inflow rates and efficiencies.

\item
Each time a gas inflow is stalled, one may expect star formation to be triggered.
It is therefore important to discriminate between `nuclear' star formation on scales of $<50$\,pc which may directly influence further gas inflow, and serendipitous `circumnuclear' star formation on scales of 0.1--2\,kpc that does not obviously do so.

\item
There is no overwhelming evidence for active young ($<50$\,Myr) starbursts in the tens of parsecs around AGN. In contrast, it appears there is a delay of $\sim100$\,Myr between a starburst and the onset of accretion to an AGN.
This may be because it is only during phases when slow stellar winds predominate that gas can be accreted to small scales.
Such a scenario points to a direct link between the properties of star-forming clusters on 10--50\,pc scales and the compact structures observed on 0.1--1\,pc scales.
It would also imply that the torus is a complex dynamical entity, the properties of which are coupled to nuclear star formation.

\item
In answer to the question posed in the title of this contribution, it is now becoming clear that stellar winds {\em do} play a decisive role in feeding Seyferts.
But it is also clear that other processes are also important.

\end{itemize}



\end{document}